\input harvmac
%
%
\message{S-Tables Macro v1.0, ACS, TAMU (RANHELP@VENUS.TAMU.EDU)}
%
%
\newhelp\stablestylehelp{You must choose a style between 0 and 3.}%
\newhelp\stablelinehelp{You 
should not use special hrules when stretching
a table.}%
\newhelp\stablesmultiplehelp{You have tried to place an S-Table  
inside another
S-Table.  I would recommend not going on.}%
%
%
\newdimen\stablesthinline
\stablesthinline=0.4pt
\newdimen\stablesthickline
\stablesthickline=1pt
%
%
\newif\ifstablesborderthin
\stablesborderthinfalse
\newif\ifstablesinternalthin
\stablesinternalthintrue
\newif\ifstablesomit
\newif\ifstablemode
\newif\ifstablesright
\stablesrightfalse
%
%
\newdimen\stablesbaselineskip
\newdimen\stableslineskip
\newdimen\stableslineskiplimit
%
%
\newcount\stablesmode
\newcount\stableslines
\newcount\stablestemp
\stablestemp=3
\newcount\stablescount
\stablescount=0
\newcount\stableslinet
\stableslinet=0
%
%
%
\newcount\stablestyle
\stablestyle=0
%
%
\def\stablesleft{\quad\hfil}%
\def\stablesright{\hfil\quad}%
%
%
\catcode`\|=\active%
%
%
\newcount\stablestrutsize
\newbox\stablestrutbox
\setbox\stablestrutbox=\hbox{\vrule height10pt depth5pt width0pt}
\def\stablestrut{\relax\ifmmode%
                         \copy\stablestrutbox%
                       \else%
                         \unhcopy\stablestrutbox%
                       \fi}%
%
%
\newdimen\stablesborderwidth
\newdimen\stablesinternalwidth
\newdimen\stablesdummy
\newcount\stablesdummyc
\newif\ifstablesin
\stablesinfalse
%
%
\def\begintable{\stablestart%
  \stablemodetrue%
  \stablesadj%
  \halign%
  \stablesdef}%
\def\stablesadj{%
  \ifcase\stablestyle%
    \hbox to \hsize\bgroup\hss\vbox\bgroup%
  \or%
    \hbox to \hsize\bgroup\vbox\bgroup%
  \or%
    \hbox to \hsize\bgroup\hss\vbox\bgroup%
  \or%
    \hbox\bgroup\vbox\bgroup%
  \else%
    \errhelp=\stablestylehelp%
    \errmessage{Invalid style selected, using default}%
    \hbox to \hsize\bgroup\hss\vbox\bgroup%
  \fi}%
\def\stablesend{\egroup%
  \ifcase\stablestyle%
    \hss\egroup%
  \or%
    \hss\egroup%
  \or%
    \egroup%
  \or%
    \egroup%
  \else%
    \hss\egroup%
  \fi}%
\def\stablestart{%
  \ifstablesin%
    \errhelp=\stablesmultiplehelp%
    \errmessage{An S-Table cannot be placed within an S-Table!}%
  \fi
  \global\stablesintrue%
  \global\advance\stablescount by 1%
  \message{<S-Tables Generating Table \number\stablescount}%
  \begingroup%
  \stablestrutsize=\ht\stablestrutbox%
  \advance\stablestrutsize by \dp\stablestrutbox%
  \ifstablesborderthin%
    \stablesborderwidth=\stablesthinline%
  \else%
    \stablesborderwidth=\stablesthickline%
  \fi%
  \ifstablesinternalthin%
    \stablesinternalwidth=\stablesthinline%
  \else%
    \stablesinternalwidth=\stablesthickline%
  \fi%
  \tabskip=0pt%
  \stablesbaselineskip=\baselineskip%
  \stableslineskip=\lineskip%
  \stableslineskiplimit=\lineskiplimit%
  \offinterlineskip%
  \def\borderrule{\vrule width \stablesborderwidth}%
  \def\internalrule{\vrule width \stablesinternalwidth}%
  \def\thinline{\noalign{\hrule height \stablesthinline}}%
  \def\thickline{\noalign{\hrule height \stablesthickline}}%
  \def\trule{\omit\leaders\hrule height \stablesthinline\hfill}%
  \def\ttrule{\omit\leaders\hrule height \stablesthickline\hfill}%
  \def\tttrule##1{\omit\leaders\hrule height ##1\hfill}%
  \def\stablesel{&\omit\global\stablesmode=0%
    \global\advance\stableslines by 1\borderrule\hfil\cr}%
  \def\el{\stablesel&}%
  \def\elt{\stablesel\thinline&}%
  \def\eltt{\stablesel\thickline&}%
  \def\elttt##1{\stablesel\noalign{\hrule height ##1}&}%
  \def\elspec{&\omit\hfil\borderrule\cr\omit\borderrule&%
              \ifstablemode%
              \else%
                \errhelp=\stablelinehelp%
                \errmessage{Special ruling will not display properly}%
              \fi}%
  \def\stmultispan##1{\mscount=##1 \loop\ifnum\mscount>3
\stspan\repeat}%
  \def\stspan{\span\omit \advance\mscount by -1}%
  \def\multicolumn##1{\omit\multiply\stablestemp by ##1%
     \stmultispan{\stablestemp}%
     \advance\stablesmode by ##1%
     \advance\stablesmode by -1%
     \stablestemp=3}%
  \def\multirow##1{\stablesdummyc=##1\parindent=0pt\setbox0\hbox\bgroup%
    \aftergroup\emultirow\let\temp=}
  \def\emultirow{\setbox1\vbox to\stablesdummyc\stablestrutsize%
    {\hsize\wd0\vfil\box0\vfil}%
    \ht1=\ht\stablestrutbox%
    \dp1=\dp\stablestrutbox%
    \box1}%
  
\def\stpar##1{\vtop\bgroup\hsize ##1%
     \baselineskip=\stablesbaselineskip%
     \lineskip=\stableslineskip%
      
\lineskiplimit=\stableslineskiplimit\bgroup\aftergroup\estpar\let\temp=}%
  \def\estpar{\vskip 6pt\egroup}%
  \def\stparrow##1##2{\stablesdummy=##2%
     \setbox0=\vtop to ##1\stablestrutsize\bgroup%
     \hsize\stablesdummy%
     \baselineskip=\stablesbaselineskip%
     \lineskip=\stableslineskip%
     \lineskiplimit=\stableslineskiplimit%
     \bgroup\vfil\aftergroup\estparrow%
     \let\temp=}%
  \def\estparrow{\vfil\egroup%
     \ht0=\ht\stablestrutbox%
     \dp0=\dp\stablestrutbox%
     \wd0=\stablesdummy%
     \box0}%
  \def|{\global\advance\stablesmode by 1&&&}%
  \def\|{\global\advance\stablesmode by 1&\omit\vrule width 0pt%
         \hfil&&}%
  \def\vt{\global\advance\stablesmode by 1&\omit\vrule width  
\stablesthinline%
          \hfil&&}%
  \def\vtt{\global\advance\stablesmode by 1&\omit\vrule width
\stablesthickline%
          \hfil&&}%
  \def\vttt##1{\global\advance\stablesmode by 1&\omit\vrule width ##1%
          \hfil&&}%
  \def\vtr{\global\advance\stablesmode by 1&\omit\hfil\vrule width%
           \stablesthinline&&}%
  \def\vttr{\global\advance\stablesmode by 1&\omit\hfil\vrule width%
            \stablesthickline&&}%
  \def\vtttr##1{\global\advance\stablesmode by 1&\omit\hfil\vrule  
width ##1&&}%
  \stableslines=0%
  \stablesomitfalse}
\def\stablesdef{\bgroup\stablestrut\borderrule##\tabskip=0pt plus 1fil%
  &\stablesleft##\stablesright%
  &##\ifstablesright\hfill\fi\internalrule\ifstablesright\else\hfill\fi%
  \tabskip 0pt&&##\hfil\tabskip=0pt plus 1fil%
  &\stablesleft##\stablesright%
  &##\ifstablesright\hfill\fi\internalrule\ifstablesright\else\hfill\fi%
  \tabskip=0pt\cr%
  \ifstablesborderthin%
    \thinline%
  \else%
    \thickline%
  \fi&%
}%
\def\endtable{\advance\stableslines by 1\advance\stablesmode by 1%
   \message{- Rows: \number\stableslines, Columns:   
\number\stablesmode>}%
   \stablesel%
   \ifstablesborderthin%
     \thinline%
   \else%
     \thickline%
   \fi%
   \egroup\stablesend%
\endgroup%
\global\stablesinfalse}
%

\input epsf

\baselineskip=.55truecm
\Title{\vbox{\hbox{HUTP--96/A047}\hbox{hep-th/9610111}}}
{\vbox{\centerline{Colliding Singularities in F-theory}
\vskip .1in
\centerline{and Phase Transitions}}}
\vskip .1in
\centerline{\sl Michael Bershadsky\foot{E-mail: bershad@string.harvard.edu}
and Andrei  Johansen\foot{E-mail: johansen@string.harvard.edu}}
\vskip .2in
\centerline{\it Lyman Laboratory of Physics, Harvard University}
\centerline{\it Cambridge, MA 02138, USA}


\vskip .2in
\centerline{ABSTRACT}
\vskip .2in
We study F-theory on elliptic threefold Calabi-Yau near colliding  singularities.
We demonstrate that resolutions of those singularities generically correspond to 
transitions  to phases characterized by  new tensor multiplets and 
enhanced gauge symmetry.
These are governed by the dynamics of tensionless strings.
We also find new transition points which are associated with 
several small instantons simultaneously shrinking to zero size.
\Date{\bf October 1996}

\newsec{Introduction}

Great progress has been made recently in our understanding of  
enhanced gauge symmetries and matter contents in 
compactifications of F-theory
\ref\vafa{C. Vafa, Nucl. Phys. {\bf B 469} (1996) 403, hep-th/9602022.}
\ref\MV{D. Morrison and C. Vafa, `Compactifications of F-theory on 
Calabi-Yau Threefolds-
I, II'', hep-th/9602114, hep-th/9603161.}
\ref\kucha{M. Bershadsky, K. Intriligator,
S. Kachru, D. Morrison, V. Sadov, and C. Vafa, `Geometric 
Singularities and Enhanced
Gauge Symmetries', hep-th/9605200.} \ref\KV{S. Katz and C. Vafa,
`Matter From Geometry', hep-th/9606086.}.
In this paper we further develop the {\it geometry/physics} dictionary for some 
F-theory compactifications on  elliptic Calabi-Yau threefolds.
Such an analysis can provide us with additional evidence for string-string 
duality as well as give us a better understanding of quantum 
field theory applications of the latter
\ref\three{T. Banks, M.R. Douglas and N. Seiberg,
`Probing F-theory with Branes', hep-th/9605199\semi
N. Seiberg, `IR Dynamics on Branes and Space-Time 
Geometry', hep-th/9606017.}. 
Only elliptic Calabi-Yau manifolds appear in F-theory compactifications.  
The elliptic fibration is defined over some two-dimensional base $B$ and can be written 
 in  Weierstrass form
\eqn\wei{y^2=x^3+xf(z,w)+g(z,w)~,}
where $(z,w)$ parametrize two-dimensional base.
The appearance of singularities in elliptic fibration is responsible for the 
enhanced gauge symmetry.  
The singularities in the elliptic fiber were classified by Kodaira and are 
summarized in the table below.

\bigskip
Table 1: Kodaira Classification of Singularities
\vskip .1in
\begintable
 ${\rm ord}(f)$ | ${\rm ord}(g)$ | ${\rm ord}(\Delta)$   |
                               fiber type | singularity type  \elt
$\geq 0$ | $\geq 0$ | $0$ | smooth |none  \elt
$0$ | $0$ | $n$ |  $I_n$  | $A_{n-1}$  \elt
$\geq 1$ | $1$ | $2$| $II$ | none   \elt
$1$ | $\geq 2$ | $3$ |  $III$ | $A_1$  \elt
$\geq 2$ | $2$ | $4$ |   $IV$  | $A_2$  \elt
$2$ | $\geq 3$ | $n+6$ |  $I_n ^*$ | $D_{n+4}$  \elt
$\geq 2$ | $3$ | $n+6$ |  $I_n ^*$ | $D_{n+4}$  \elt
$\geq 3$ | $4$ | $8$ |  $IV^*$ | $E_6$  \elt
$3$ | $\geq 5$ | $9$ |  $III^*$  | $E_7$  \elt
$\geq 4$ | $5$ | $10$ |  $II^*$  | $E_8$ 
\endtable
\bigskip

Only the local structure of singularity 
around the degeneration divisor  (position of the 7-brane) is relevant.
The physical reasoning for appearance enhanced gauge symmetry is clear.
Open strings with various $(p,q)$ charges 
connecting 7-branes
\ref\witt{E. Witten, Nucl. Phys. {\bf B 460} (1996) 335.}
\vafa \
\ref\sen{A. Sen, ``F-theory and Orbifolds'',
hep-th/9605150.}  
\ref\muha{ K. Dasgupta and S. Mukhi,
`F-theory at Constant Coupling', hep-th/9606044.}
\ref\ajoh{A. Johansen, `A Comment on BPS States in 
F-theory in 8 Dimensions', HUTP--96/A030, hep-th/9608186.}  become massless
vector particles
and promote abelian symmetry to non-abelian one.
 Various singularities of F-theory compactification were analyzed in ref. \kucha\
and corresponding gauge groups have been 
found, including the non-simply laced series $B$  and
$C$ as well as $F_4$ and $G_2$. 
It appears 
that generic singularity does not usually 
correspond to the maximal gauge group associated with it  \kucha \ref\aspin{P. Aspinwall
and M. Gross, {\it The $SO(32)$ Heterotic String on K3 Surface},
hep-th/9605131.}.
The reason for this is  that there are certain monodromies along the curve
of singularities given by either internal (split  singularity) or outer (nonsplit singularity) automorphisms of the root lattice. 
It has also been found that the gauge groups associated with two intersecting 7-branes cannot have simultaneously a perturbative explanation from the heterotic point of view.
In the usual setup one of the gauge  groups is perturbative while the other one is realized
via small instantons shrinking to zero size on the heterotic side.
We are going to call  the two intersecting D-branes with the gauge groups as
``collision of the gauge groups'', borrowing this term from mathematics 
(``collision of singularities''). 
This  theory is nonanomalous  only if the intersecting 7-branes correspond to
 either two 
$SU$'s or $SU$ and $SO$  gauge groups.  One can analyse these types of
collisions of singularities following the  Katz-Vafa suggestion \KV.
The basic idea is very simple. Fibering 8-dimensional theory with gauge symmetry ${\cal G}$,
one obtains 6-dimensional compactification. The fibration parameter $t$  can be interpreted as
the  
$vev$ of  the adjoint scalar.   
For $t \neq 0$ the theory possesses  $G \times G'  \times U(1) \subset  {\cal G}$
symmetry. The spectrum of 6-dimensional theory follows unambiguously
from the Higgs mechanism. 
 This method works only if one can break ${\cal G} \longrightarrow G \times G' \times U(1) $
by giving $vev$ to adjoint matter. 
Indeed,  groups $SU(n) \times SU(m) \times U(1)$ or $SU(n) \times SO(2m) \times U(1) $ can be obtained by  breaking down $SU(n+m)$ or  $SO(2m+2n)$, respectively.  It is
easy to explain this process as splitting a Dynkin diagram into two parts by
removing one node (which corresponds to $U(1)$ factor).  Therefore, the collisions of 
gauge groups
whose Dynkin diagrams cannot be embedded into bigger one cannot be analyzed in this way.

The simplest example where one cannot find the spectrum of  the theory is 
the collision of  $SO(n)$ groups.  It is impossible to satisfy the 
anomaly cancellation conditions for F-theory. 
It seems that the local field theory description does not exist.

To be specific, suppose that 
a discriminant locus contains two intersecting D-branes, say $D$ 
and $D'$, each corresponding to the gauge groups $SO(2n+8)$.
From the mathematical point of view, the  Calabi-Yau threefold with two colliding  $I^* _n$ singularities
($I^* _n$ singularity corresponds to $SO(2n+8)$) appears to be very singular and requires a resolution. 
Resolving the singularity it is not enough to blow up a fiber and one needs also to blow up a base at the point of intersection.  
That means replacing an intersection point by the whole $P^1$ and, as the result, the divisors $D$ and $D'$ do not intersect each other on the blown up surface.  
 It happens very often that
the new sphere becomes also a component of the discriminant locus. In this situation one may get a nonperturbative gauge symmetry enhancement. 
The new gauge group $G$ is associated with the
blown up divisor and is determined by the structure of elliptic fibration. 
It appears that after the resolution one can also satisfy the anomaly factorization conditions.

From the physical point of view this collision can be understood as follows.
One can start  with perfectly well defined theory with two intersecting D-branes
carrying, say for definiteness, gauge groups $SO(\ast) \times SU(\ast)$.  By adjusting the 
$vev$'s
 of the hypermultiplets one can enhance the second group to $SO(\ast)$, keeping the first
one intact.  At this very moment we reach the transition point
and the local field theory description becomes inconsistent. 
A new branch with an additional tensor multiplet, known as ``Coulomb''  branch
\ref\seiwitt{N. Seiberg and E. Witten, Nucl. Phys. {\bf B 471} (1996) 121; hep-th/9603003.}, 
is attached at the transition point.
It is characterized by the vacuum expectation value of the scalar field in the tensor multiplet.
This branch becomes the standard Coulomb branch after compactification 
down to 4-dimensions.
The transition point is governed by the 
dynamics of the tensionless strings \seiwitt  \ref\vafaetal{M.R. Douglas, S. Katz and C. Vafa,
`Small Instantons, Del Pezzo Surfaces and type I' Theory', hep-th/9609071.}
\ref\klemm{A. Klemm, P. Mayr and C. Vafa, `BPS States of Exceptional Noncritical Strings', hep-th/9607139.}
\ref\Wilast{E. Witten, `Physical Interpretation of Certain Strong Coupling 
Singularities', IASSNS-HEP-96/97, hep-th/9609159.}
\ref\Sieb{N. Seiberg, `Non-trivial Fixed Points of the Renormalization Group in
Six Dimensions', RU-96-85, hep-th/9609161.}.
The appearance of tensionless strings is clear if we approach the transition point from the 
Coulomb branch.   
The tensionless strings correspond to 3-branes wrapped around the vanishing 
(blown up or down) 2-cycle.
The tension of the strings is given by the $vev$ of tensor multiplet, or
expressing in mathematical terms, by the area of the blown up sphere. 
It is also worth mentioning that after the phase transition, the theory ceases to have 
heterotic dual. 

In the Coulomb branch, we get a perfectly nonanomalous theory.
In some cases the gauge symmetry gets enhanced.  
The gauge coupling constant of this new nonperturbative symmetry is
governed by the $vev$ of the tensor multiplet
\eqn\area{ {1 \over g^2} \sim vev \sim {\rm area}(P^1)~.} 
Approaching the transition point from the Coulomb 
branch one recovers a strong coupling transition. 
This also requires a fine 
tuning of both hypermultiplets and tensor multiplets.
It is worth mentioning that passing through the transition point to the Coulomb branch
makes some of the original hypermultiplets heavy.
In what follows we will be able to analyze various collisions and to
predict corresponding gauge groups and their  matter contents. 

Some of the phase transitions discussed in this paper are known to be 
associated to a single small $E_8$ instanton. 
We also find new phase transition points
related to two, three and five small $E_8$ instantons collapsing simultaneously.

\bigskip

Organization of this paper is as follows: we first review the anomaly factorization condition
for F-theory compactifications.  
In section 3 we present mathematical discussion of the blowup procedure. 
In section 4 we discuss the physical interpretations of the 
collisions using the  blowup technique.  
Finally, we conclude with an analysis of  the role of small $E_8$ instantons and 
with the discussion of
new phenomena whose interpretation  in  terms of tensionless strings is unclear.

\newsec{Anomaly cancellation and colliding singularities}

Here we will review the conditions necessary for anomaly cancellation in 6-dimensions. Anomaly cancelation via Green-Schwarz mechanism \ref\GS{M.B. Green and J.H. Schwarz, Phys. Lett. 
{\bf 149 B} (1984) 117.} \ref\aug{A. Sagnotti, Phys. Lett. {\bf 294 B} (1992) 196, hep-th/9210127.}
requires that a certain  $8$-form should be factorized. 
This implies that  at least  the coefficient in front of
${\rm tr} R^4$ should vanish, which is equivalent to some relation 
between the number of
$n_V$ (vector), $n_H$ (hyper) and $n_T$ (tensor) multiplets:
\eqn\ten{n_H-n_V+29 n_T=273~.}
The resolution of the singularities changes the number of tensor multiplets. 
Only the local structure of the singularity is relevant and therefore this implies $\delta n_H- \delta n_V+29=0$  for each  blowup. 

There are other conditions that also should be satisfied in order for 
factorization to take place. 
The conditions for anomaly cancellation in F-theory  has been found in ref. 
\ref\sadov{V. Sadov, `Generalized Green-Schwarz Mechanism in F Theory',
IASSNS-HEP-96/58, hep-th/9606008.}.
Let us denote the irreducible components of the discriminant locus as
$D_a \subset B$.  Each component of the discriminant locus $D_a$ gives rise to the 
gauge group $G_a$,
depending on the singularity in the elliptic fibration. The gauge fields propagate inside 
7-branes wrapped around divisors $D_a$.

The anomaly factorization conditions
are
\eqn\anomaly{\eqalign{
\sum_{(R_a,R_b)} n_{(R_a,R_b)} {\rm Ind }(R_a) {\rm Ind }(R_b)=(D_a \cdot D_b), \cr
{\rm Ind} (Ad_a)-\sum_R {\rm Ind}(R_a)~n_{R_a}=6(K\cdot D_a),\cr
y_{Ad_a}-\sum_R y_{R_a}~n_{R_a}=-3(D_a\cdot D_a),\cr 
x_{Ad_a}-\sum_R x_{R_a}~n_{R_a}=0~.}}
Here, ${\rm Ind} (R)$ stands for the Dynkin index of representation $R$,
the numbers $n_{R_a}$ denote a multiplicity of the representation
$R_a$ of the gauge group $G_a$, and $n_{(R_a, R_b)}$ denotes
 the number of mixed representations. 
Parameters $x_R$ and $y_R$ are defined by the following decomposition
\eqn\decomp{{\rm tr}_R F^4=x_R {\rm tr} F^4 + y_R ({\rm tr} F^2)^2~,}
assuming $R$ has two independent order four invariants;
${\rm tr}_R$ denotes the trace in the representation $R,$ and ${\rm tr}$ stands for the trace in
some standard representation (usually fundamental representation).

Let us explain how the local field theory ``feels'' the collision of singularities
and why one should resolve it.
Suppose that discriminant locus contains two intersecting D-branes, say $D$ 
and $D'$, each corresponding to the gauge groups $G$ and $G'$.
The local analysis of D-branes \ref\bsv{M. Bershadsky, V. Sadov and C. Vafa,
 Nucl. Phys. {\bf B 463} (1996) 398. } \sadov\ 
implies that each intersection point should give rise to hypermultiplets in mixed representation 
$(R,R')$.  
Moreover, one can discuss the anomaly cancellation locally at each intersection point
(the first equation in  \anomaly).
The condition for
having a consistent   theory reads that  either  $ n_{(R,R')} ={\rm Ind }(R) ={\rm Ind }(R')=1$
or  $n_{(R,R' )}=1/2$,  and 
indices ${\rm Ind} (R)=1$ and  ${\rm Ind} (R')=2$  (the latter solution is possible only
for pseudoreal representation $(R,R')$). Now we can see that the case $SO(n) \times SO(m)$ is completely ruled out because all indices  $ {\rm Ind}  \geq 2$ for $n,m \geq 7$.

\newsec{Blowups}

One of the main tools in resolution of singularities is the blowing up procedure
\ref\hart{R. Hartshorne, {\it Algebraic Geometry}, Springer.}.
We will discuss here a blowing up of an algebraic surface at a point $P$.
We parametrize an open neighborhood $U_P \subset B$ using affine coordinates
$(z,w)$. One can think about $U_P$ being embedded in $A^2$ (affine two-dimensional space).
Consider the product $A^2 \times P^1$, 
which is a quasi projective variety, with $(y_1,y_2)$ being homogeneous
coordinates in $P^1$. 
Define the{ \it  blowing up} of $ B$
at point $P=(0,0)$ as the closed subset  in 
 $\hat U_P \subset A^2 \times P^1$, defined by the 
equations 
\eqn\bl{z y_2=w y_1~.}
We will denote the blown up surface by $\hat B$.

Let us denote the map $\pi :  U_P  \times P^1 \rightarrow  U_P$, defined by the
above equation. Then  $\pi^{-1}(Q)$ for $Q \neq P$ consists of one point, while 
$\pi^{-1}(P)=P^1$.  In other words the blowup procedure ``replaces''
the point $P=(0,0)$ by $P^1$. This $P^1$ is called an exceptional divisor
and we reserve the notation $E$  for it.
To parameterize the blown up base around the point $P$ one can use either
$(z, \xi=y_2/y_1)$ or $(w, \eta=y_1/y_2)$. 
The map $\pi_{\ast}: H^2 (\hat B, {\bf Z}) \rightarrow H^2 (B, {\bf Z})$ respects the intersection form and has a kernel generated by $E$.  
Denote by $D^*$ the full preimage of a divisor $D$ in $\hat B$.
Suppose that the divisor $D$ passes through the point $P$ we are blowing up.
In this case the divisor $D^*$ happens to be reducible.
Namely, $D^*=\hat D+ E$, where both $\hat D$ and $E$ are irreducible.
Suppose two  divisors   $D$ and $D'$  intersect  each other at the  point $P$. 
In the blown up surface $\hat B$ the irreducible divisors $\hat D$ and ${\hat D}'$
do not intersect each other
\eqn\int{ \eqalign{
&(\hat D \cdot {\hat D}')=\big( (D^* -E)  \cdot  ({D'}^* -E) \big)=0 \cr
&(\hat D \cdot E)=1~,~~( {\hat D}' \cdot E)=1 ~.\cr
}}
Intersection pairing of the divisors that do not pass through the
intersection point remains unchanged.
The canonical class of the blown up surface is equal to
$\hat K=K^* +E$.

The elliptic fibration can be pulled up on the 
blown up surface $\hat B$. 
The new fibration is defined as
\eqn\blfib{\eqalign{
f(z,w) \longrightarrow  f(z,\xi)&=f(z,z \xi)\cr
g(z,w) \longrightarrow g(z,\xi)&=g(z,z \xi) ~.\cr}}
For our purpose having elliptic fibration is not enough. 
One has to check whether the elliptic fibration on the blown up surface determines the
Calabi-Yau manifold.
The condition for having  Calabi-Yau is that  canonical class
can be written in terms of the irreducible components of the discriminant locus 
\eqn\can{K(B)=- \sum _{i} a_i  [D_i ]-a[D] -a' [D']~,  }
where we explicitly singled out two intersecting components of the discriminant locus.
Upon the blowup the divisor $E$ becomes a new component of the discriminant.
The relation for the blown up surface reads
\eqn\cantwo{K^*(B)+[E]=- \sum _{i} a_i  [D^*_i ]-a[\hat D] -a' [\hat D']-b[E]~,  }
where $b$ is determined by the type of singularity of the fiber over the blown up sphere. 
In order for these two relations to be satisfied it is necessary that 
\eqn\CY{a+a'-b=1~.}
This relation severely constrains possible collisions, which admit resolutions by blowing up 
the base and preserve the Calabi-Yau condition \CY. 
The coefficients $a_i$ are summarized in the table below.

\bigskip
Table 2. Coefficients $a_i$.
\vskip .1in
\begintable
 none | $I_n$ | $II$ | $III$ | $IV$ | $I^* _n$ | $IV^*$ | $III^*$ | $II^*$ \elt
 0 | ${n \over 12}$ | ${1 \over 6}$ | ${1 \over 4}$ | ${1 \over 3}$ | ${1 \over 2} + {n \over 12}$ | ${2 \over 3}$ | ${3 \over 4}$ | ${5 \over 6}$ 
\endtable
\bigskip

It is worth mentioning that there are several collisions, known as dual, for which $a+a'=1,~b=0$. These collisions do not 
lead to any enhanced gauge symmetry.

To make blowup one has to adjust an appropriate number of parameters. Blowup does not introduce any new complex parameters. 
To be able to pull up the elliptic fibration on the blown up base
the Calabi-Yau manifold should be described by the following Weierstrass form \MV
\eqn\WST{ y^2=x^3+x\sum_{l+k \geq 4} z^lw^k f_{l,k}+\sum_{l+k \geq 6} z^lw^k g_{l,k}.}
We assumed that the base of the elliptic fibrartion is Hirzebruch surface.
Condition \CY\ appears to be equivalent to the restrictions on the 
coefficients in the expansion \WST.
The blowup point $(z,w)=(0,0)$ lies on the discriminant locus, 
which in general may correspond to a very mild singularity, say $I_1$.
This singularity does not lead to any  gauge symmetry enhancement.
We mostly, are going to be interested in different situations, when the  
blowup point  coincides with  the intersection  of two divisors, 
each corresponding to
nonabelian gauge symmetry.
We will return back to this discussion in the last section.

\newsec{The cases}

As we discussed in the introduction, collisions of singularities lead to 
singular Calabi-Yau manifolds. 
To resolve these singularities, one needs to make an appropriate number of blowups.
In this section we discuss only those collisions which can be blown up without violation
of Calabi-Yau condition \CY.
The allowed collisions of singularities can be classified by the value 
of the modular invariant function $J(\tau)$ 
at the collision point 
\ref\miranda{R. Miranda, {\it Smooth Models for 
Elliptic Threefolds} in R. Friedman and D.R. Morrison, editors, ``The Birational 
Geometry of Degenerations'', Birkhauser, 1983.}.  
Only the fibers with the same value of $J(\tau)$ at the intersection
point can collide. 
Some ``collisions'' of singular elliptic curves produce singular 
Calabi-Yau threefolds 
that can be resolved by blowing up a fiber (small resolutions). For example, 
the collisions $I_n \times I_m$ ($SU(\ast) \times SU(\ast)$)
and $I_n \times I_m ^*$  ($SU(\ast) \times SO(\ast)$) are exactly of 
this type (considered  in  \kucha \KV).

Other collisions do not have 
 one-dimensional resolutions. 
These situations can be resolved by suitable blowing up the base at the collision point
and pulling up the fibration to the blown up surface. 
In some cases one has to repeat the blowup 
operation several times. In  this process one creates more collisions.
The whole procedure stops when all collisions allow small resolutions and can be resolved by blowing up a fiber.

At this point it is instructive to compare the physical interpretation of 
``blowing up''  a fiber with that of  ``blowing up'' a base. 
The blowup of a fiber becomes visible upon compactification down to 5 dimensions.
The area of the blown up sphere coincides with the $vev$ of the scalar field, which
parametrizes the Coulomb branch in 5 dimensions. The blowup of the base
is already visible in 6 dimensions and corresponds to 
the appearance of  a new tensor multiplet.

In the examples considered below,  we usually end up
with enhanced gauge symmetry, say $G_1 \times {\cal H} \times G_2$,
upon the resolution. The gauge groups 
$G_1$ and $G_2$ correspond to colliding singularities and in some sense serve as our 
``initial data''.  The intermediate factor ${\cal H}$ describes the gauge symmetry enhancement.
The matter content  is given by 
hypermultiplets in various representations. 
The part  of the hypermultiplet spectrum, charged only with respect to either
$G_1$ or $G_2$, depends on particular details of the theory, such as
the choice of the base of the elliptic fibration and the choice of the divisors.
The matter in ${\cal H}$ representation as well 
as in the mixed representations of 
$G_{1,2} \times {\cal H}$ are {\it universal} and depend only on the local structure of the colliding singularities.

In order to be specific and make the examples more transparent,
we need to specify some details of the theory. 
For simplicity we assume that intersecting divisors are homologically spheres and
their  intersections are: $D \cdot D=n_1$, $D' \cdot D'=n_2$ and $D \cdot D'=1$.
After the blowup the irreducible components $\hat D$ and $\hat D'$ have the following intersection pairing
\eqn\inner{\eqalign{{\hat D}\cdot {\hat D}=n_1-1,~~~{\hat D}'\cdot {\hat D}' =n_2-1,~~~E\cdot E=-1,~~\cr
{\hat K}\cdot {\hat D}=-n_1-1,~~~{\hat K}\cdot {\hat D}'=-n_2-1, ~~~{\hat K}\cdot E=-1~~.}}
For example, in the case of Hirzebruch surface ${\bf F}_n$, if  the singularities are along
$D=D_u$ (the base) and $D'=D_s$ (the $P^1$ fiber), then $n_1,~n_2$ are equal to $n$ and $0$
or vice versa.

\subsec{Dual singularities}

The case $I^*_0\times I^*_0$ is the  first example of the collision of dual fibers. It is also special because  it can be realized for any value of $J$.
Without the second singularity each of these degenerations gives rise to the consistent theory 
with maximal allowed gauge group $SO(8)$ (which corresponds to split singularity),
each having
$(n_i+4)({\bf 8_v+8_c+8_s})$ matter hypermultiplets.
As it was noticed in \kucha, when the singularities collide, one can not get a 
consistent local field theory description.

To resolve such a singularity 
one needs to perform a blowup of an intersection  point $P$.  
 For dual singularities the exceptional divisor $E$ {\it is not}  a component of the 
discriminant locus. 
The fiber over $E$ is smooth and there is no gauge symmetry enhancement.
So the resolution of the paradox is, that on the blown up base, 
the divisors $\hat D$ and $\hat D'$
do not meet each other and there are no mixed representations. 
Going through the analysis of the 
anomaly equations \anomaly\ or codimensional counting, 
we conclude that  in the 
new branch (with non zero $vev$ of the tensor multiplet) there is still a
gauge group $SO(8) \times SO(8)$ with $(n_i+3)({\bf 8_v+8_c+8_s})$ matter hypermultiplets
for each factor
\foot{ This result is consistent with the examples discussed in refs. \MV  \ \ref\blum{J.D. Blum
and A. Zaffaroni, `An Orientifold from F Theory', hep-th/9607019\semi
J.D. Blum, `F Theory Orientifolds, M Theory Orientifolds, and
Twisted Strings', hep-th/9608053.} 
\ref\ferrara{S. Ferrara, R. Minasian and A. Sagnotti, `Low-energy analysis of M and F theories
in Calabi-Yau threefolds', hep-th/9604097 \semi 
C. Angelantonj, M. Bianchi, G. Pradisi, A. Sagnotti and Y.S. Stanev,
`Comments on Gepner Models and Type I Vacua in String Theory', hep-th/9607229.}
in the case of some orientifold compactifications.}. 
The change in the matter spectrum $(n_i+4) \rightarrow (n_i+3)$ can be
interpreted as  the result of some coupling between the matter hypermultiplets and the
tensor multiplet (some hypermultiplets become heavy).

In fact there is nothing special about $SO(8)$ gauge group. 
It corresponds to 
$I^{*s} _0$ singularities. 
We can also discuss $SO(7)$ ($I^{*ss} _0$)
or $G_2$ ($I^{*ns}_0$) singularity. Indeed, by giving expectation value to
two spinors, we can higgs $SO(8)$ down to $G_2$. 
The case of
$G_2$ is interesting: it corresponds to nonsplit singularity and one can immediately
count the dimension of the moduli space of hypermultiplets.  
The collision $I^{*ns}_0    \times I^{*ns}_0$ corresponds to
$G_2 \times G_2 $ gauge group with the matter contents $(3n_i +7)~{\bf 7}$ without any 
{\it mixed} matter.

For simplicity we present the parameter counting for $P^1 \times P^1$ being the base ($n_1=n_2=0$), but this computation  can be generalized for any Hirzebruch surface. It follows from the structure of polynomials $f,g$,
that the number of independent parameters is $7^2+10^2-4=145$
\foot{$4=6-2$ comes from $SL(2,{\bf Z})\times SL(2,{\bf Z})$ residual symmetry.
Each $P^1$ has a marked point (location of singularity), which explains $-2$.}.
The total number of $n_H-n_V=273-29 n_T$. 
After the blowup $n_T=2$ and, as a  result, we get a perfect match
\eqn\count{145+7 \cdot 7 + 7 \cdot 7 -14 -14=215.}
The counting for other cases $I^*_0 \times I^*_0$ is obvious because they are related  to 
$G_2 \times G_2$ by Higgs mechanism.

There are several cases like the above one, namely, $II \times II^*$, $III \times III^*$ and $IV \times IV^*$. 
In all these cases the exceptional divisor $E$ does not belong to the discriminant locus and
therefore there is no gauge symmetry enhancement. The discussion is very similar to the $I^* _0 \times 
I^* _0$ collision.
For colliding singularities these gauge theories are anomalous because they should necessarily have matter in the mixed representations.  For the blown up base the divisors 
$\hat D$ and $\hat D'$ do not intersect and therefore in this phase the theory does not have 
any mixed representations and one can satisfy the anomaly factorization conditions.
We summarize all these results in the table below.

\bigskip
\vbox{Table 3. Gauge groups and matter for collisions of dual singularities.
\vskip .1in
\begintable
Singularity | Gauge group | Matter  \elt
$I^{*ns} _0 \times I^{*ns} _0$ | $G_2 \times G_2$ | $(3 n_1+7) ({\bf  7,1})+ (3 n_2+7) ({\bf 1, 7})$ \elt
 $II \times II^*$ | $E_7$ | ${1 \over 2}(n_1+7) {\bf  56}$ \elt
 $III \times III^*$ | $SU(2) \times E_7$ | ${1 \over 2}(n_1+7) {\bf  (1,56)} + (6 n_2+10) {\bf (2,1)}$ \elt
 $IV^s \times IV^{*ns}$  | $SU(3) \times F_4$ | $(n_1+4) {\bf (1,26)}+ (6n_2 +12){\bf (3,1)}$ 
\endtable}

\bigskip

\subsec{ $J=\infty$ collisions}

We first consider the collisions of  $I^* _n \times I^* _m$ ($SO(\ast) \times SO(\ast)$) singularities with $n+m >0$.
In all these cases one has to blow up the base only once and the exceptional divisor $E$
becomes the component of the {\it discriminant locus}.
Therefore, we should expect the enhancement of gauge symmetry. 
The fiber over $E$ generically corresponds to $I_{m+n}$  ($A_{n+m-1}$) 
singularity and jumps to 
an $I^*_m$ ($I^*_n$) singularity at the points of intersection with the $D$ and $D'$ 
divisors.
We claim that for $n+m$ even the singularity along $E$ is nonsplit and it corresponds to
$Sp((n+m)/2)$ gauge symmetry\foot{In principle we have to choose between $SU(n+m)$
and $Sp((n+m)/2)$, which is the choice between split and nonsplit cases. However, the
split case does not satisfy the anomaly factorization condition.}. There are three different cases to consider depending on whether the colliding singularities are split or nonsplit.
All these cases differ from each other in the matter content.
We summarized the results for case $n+m$ being even in the table below

\bigskip

Table 4. Gauge groups and universal matter for $J=\infty$ collisions for even $m+n$.
\vskip .1in
\begintable
Singularity| Group, Universal Matter \elt
\vbox{\vskip .05in
\hbox{$I^{s*}_n \times I^{s*}_m$}
\vskip .08in} |
\vbox{\vskip .05in
\hbox{ $SO(2n+8) \times Sp((n+m)/2) \times SO(2m+8)$}
\vskip .05in
\hbox{${1 \over 2}({\bf 2n+8,n+m,1})+{1 \over 2}({\bf 1,n+m,2m+8})$ }
\vskip .03in
}\elt
\vbox{\vskip .05in
\hbox{$I^{ns*}_n \times I^{s*}_m$}
\vskip .08in} |
\vbox{\vskip .05in
\hbox{ $SO(2n+7) \times  Sp((n+m)/2) \times SO(2m+8)$}
\vskip .05in
\hbox{${1 \over 2}({\bf 2n+7,n+m,1})+{1 \over 2}({\bf 1,n+m,1})+
{1 \over 2}({\bf 1,n+m,2m+8})$}
\vskip .03in}\elt
\vbox{\vskip .05in
\hbox{$I^{ns*}_n \times I^{ns*}_m$}
\vskip .08in} |
\vbox{\vskip .05in
\hbox{ $SO(2n+7) \times Sp((n+m)/2) \times SO(2m+7)$}
\vskip .05in
\hbox{${1 \over 2}({\bf 2n+7,n+m,1})+({\bf 1,n+m,1})+{1 \over 2}({\bf 1,n+m,2m+7})$ }
\vskip .03in}
\endtable

\bigskip 

It is remarkable, that  in all these cases one can not break 
$Sp((n+m)/2)$ by giving expectation value to $({\bf1,4,1})$.  We present  here only the {\it universal} part of the matter spectrum, which depends on the local structure of the 
singularity. 

From the mathematical point of view the case of $(n+m)$ odd is unclear. The group $SU(n+m)$ does not satisfy the 
anomaly conditions \anomaly, while  the nonsplit $I_{2k+1}$ singularities
does not seem to correspond to any gauge groups (at least the gauge groups are 
unknown \kucha).
Nevertheless, one can make some predictions based on 
the Higgs mechanism. 
It   follows from the Higgs mechanism that in case of $n+m$ 
being odd the gauge group  also gets enhanced by a factor $Sp((n+m)/2)$. 
The conjectured answers for $n+m$ being 
odd are summarized in the table below

\bigskip

\vbox{Table 5. Gauge groups and universal matter for $J=\infty$ collisions for odd $m+n$.
\vskip .1in
\begintable
Singularity| Group, Universal Matter \elt
\vbox{\vskip .05in
\hbox{$I^{s*}_n \times I^{s*}_m$}
\vskip .08in} |
\vbox{\vskip .05in
\hbox{ $SO(2n+8) \times Sp((n+m+1)/2) \times SO(2m+8)$}
\vskip .05in
\hbox{${1 \over 2}({\bf 2n+8,n+m+1,1})+({\bf 1,n+m+1,1})+{1 \over 2}({\bf 1,n+m+1,2m+8})$}
\vskip .03in}\elt
\vbox{\vskip .05in
\hbox{$I^{ns*}_n \times I^{s*}_m$}
\vskip .08in} |
\vbox{\vskip .05in
\hbox{ $SO(2n+7) \times  Sp((n+m+1)/2) \times SO(2m+8)$}
\vskip .05in
\hbox{${1 \over 2}({\bf 2n+7,n+m+1,1})+{3 \over 2}({\bf 1,n+m+1,1})+{1 \over 2}({\bf 1,n+m+1,2m+8})$ }
\vskip .03in}\elt
\vbox{\vskip .05in
\hbox{$I^{ns*}_n \times I^{ns*}_m$}
\vskip .08in} |
\vbox{\vskip .05in
\hbox{ $SO(2n+7) \times Sp((n+m-1)/2) \times SO(2m+7)$}
\vskip .05in
\hbox{${1 \over 2}({\bf 2n+7,n+m-1,1})+{1 \over 2}({\bf 1,n+m-1,2m+7})$ }
\vskip .03in}
\endtable }

\bigskip
\noindent
Singularities $I_n ^*$ with $n$ big enough can not be realized in compact 
Calabi-Yau space, because they destroy the triviality of canonical bundle.
Nevertheless, it makes perfect sense to discuss the local structure of the resolution.

The cases with $n,m \leq 12$ can be realized as collisions in compact Calabi-Yau spaces.
We start by considering the  case  $SO(12) \times SO(12)$, which corresponds to 
the collision of $I^* _2 \times I^* _2$ singularities. 
The singularity along the exceptional divisor is
$A_3$ (the type $I_4$ fiber).  One can go through tedious calculations, 
imposing the conditions found in \kucha\  and recover that 
$f(z, \xi)$ and $g(z, \xi)$ indeed corresponds to $I_4$ singularity
(we do not present these calculations here). The real issue is whether we
get $SU(4)$ or $SO(5)$. To answer this question, consider the 
leading behavior of $f(z, \xi)$ and $g(z, \xi)$  
\eqn\non{ f(z, \xi)= -3 (\xi h_0)^2 +O(z)~,~~g(z, \xi)= 2( \xi h_0)^3+O(z)}
Here, we definitely get a nonsplit singularity because the expansion for 
 $f(z, \xi)$ starts as  ${h^2 (\xi)}$ (in the split case the expansion should start as
${\tilde  h}^4 (\xi)$).  

This is consistent with the fact that we can find the
solution of anomaly equations only for $SO(5)$ gauge group.
The $SU(4)$ group requires matter in the representation
${1 \over 2}{\bf (12,4,1)}$, which does not make 
much sense. 
However this matter makes a perfect sense for $SO(5)$.
Finally, the matter spectrum is given by $(n_i+5) ~{\bf 12} +{1 \over 2} (n_i+3)~{\bf  32}$
and the {\it universal } part
\eqn\spec{
{1 \over 2}{\bf (12,4,1)}+{1 \over 2}{\bf (1,4,12)} ~. }
It is remarkable that there is no matter charged only with respect to $SO(5)$.
Now one can higgs $SO(12)$ down to $SO(n)$ for $8 \leq n <12$  in order to
``derive''  the gauge groups and the matter contents for these cases. 
In doing this we find some surprises.

Consider giving expectation value to two vectors of $SO(12)$. 
In doing this we recover the collision of two groups $SO(10) \times SO(12)$. This 
is exactly the case for which we do not have any prediction, from ref. \kucha . 
By blowing up this collision one should end up with the 
$SO(10) \times SO(5) \times SO(12)$
gauge group and matter in the  $(n_1 +3) ~({\bf 10,1,1})+(n_1+3)~ ({\bf 16,1,1})+
(n_2 +5) ~({\bf 1,1,12})+{1 \over 2}(n_2+3)~ ({\bf 1,1,32})$ representation,
as well as the {\it universal} part 
\eqn\mat{{1 \over 2}{\bf (10,4,1)}+
{\bf (1,4,1)} + {1 \over 2}{\bf (1,4,12)}   
.}
In spite of the fact that there is matter charged only with respect to nonperturbative 
$SO(5)$, one cannot higgs it. 
This case corresponds to $I_3^{ns}$ (nonsplit).
Therefore, we have a definite prediction that $I_3 ^{ns}$ should correspond to 
$SO(5)$. 
Again, it is remarkable that the group $SU(3)$ is ruled out by the same
arguments as $SU(4)$.

Going further down one can higgs the other $SO(12)$. In doing this we recover
 $SO(10) \times SO(5) \times SO(10)$ with the $(n_i+3)({\bf 10+16})$ and 
\eqn\matter{{1 \over 2}{\bf (10,4,1)}+
2 {\bf (1,4,1)}
+ {1 \over 2}{\bf (1,4,10)}   
}
hypermultiplets.
Now we have enough matter  to break nonperturbative $SO(5)$
to  $SU(2)$ without destroying both $SO(10)$.  
In this process two ${\bf (1,4,1)}$ get eaten. 
As the result  we get $(n_i+4){\bf 10}+(n_i+3){\bf 16}$ of each  $SO(10)$ 
as well as  {\it universal} mixed representations
$${1 \over 2}{\bf (1,2,10)} +{1 \over 2}{\bf (10,2,1)}~. $$  
It is remarkable that there is no  matter charged with respect to $SU(2)$.

These results can be compared with predictions coming from the collisions of singularities \miranda.
Namely, the singularity along the exceptional divisor should be of type $I_2$,
which corresponds to $SU(2)$ gauge symmetry enhancement. 
Imposing the condition for two $SO(10)$ we recover that 
\eqn\sutwo{ \eqalign{f(z, \xi)= -3 (\xi h_0)^2 +z(-6h_0h_1+f_{10}) \xi^3+O(z^2) \cr
g(z, \xi)= 2( \xi h_0)^3-z(h_0 \xi) (-6h_0h_1+f_{10}) \xi^3+O(z^2)  ~,\cr
}}
which is exactly the condition for having  $SU(2)$.
On the other hand the Higgs mechanism predicts two solutions: 
$SO(5)$ gauge group with the matter and $SU(2)$ without matter.
The resolution of this puzzle is that generically we get $SU(2)$, 
which gets  
enhanced to $SO(5)$ along some locus.

Higgsing down, one can obtain the results for $SO(9)$, $SO(8)$, $SO(7)$ or $G_2$.  
It is worth mentioning that the classification of ref. \miranda\ describes the {\it generic}
singularities.  
They correspond to the {\it minimal} possible gauge groups. For example,
in the present discussion {\it generic} $I_2$ singularity appearing upon the resolution of 
$SO(10) \times SO(10)$ collision corresponds to
$SU(2)$,  for $SO(8) \times SO(8)$ collision there is generically no gauge group enhancement.
Clearly, going to various subloci in the hypermultiplet moduli space, one can get 
different enhanced gauge symmetries.  
Namely, one can obtain $SO(8) \times SO(5) \times SO(8)$ with four matter
hypermultiplets in  ${\bf (1,4,1)}$ representation.

\subsec{$J=1$ collisions}

The only collision which can be resolved by blowing up a fiber is 
$III\times I_0^* $ ($A_1\times D_4$).

There are two cases where a blowup of  the base is required:
these are $I^*_0\times III^*$, $III^*\times III^*$ collisions. 
The gauge groups and matter content that appear for such collisions 
are given in the following table. 

\bigskip

Table 6: Gauge groups and matter content for $J=1$ collisions.
\bigskip
\begintable
 Collision | Resolution | Gauge groups  \elt
I$^*_0\times $III$^*$ |I$^*_0$,III,I$_0$,III$^*$ | $SO(8), SO(7)~{\rm or} ~G_2\times SU(2)\times E_7$\elt
III$^*\times $III$^*$ | III$^*$,I$_0$,III,I$^*_0$,III,I$_0$,III$^*$ |$E_7\times
SU(2)\times SO(7) \times SU(2)\times E_7$ \endtable
\bigskip
Here we give the minimal allowed gauge groups.

$\bullet$ $I^* _0 \times III^*$ collision ($I_0 ^* \times E_7$). 
This is the  first case where it is not enough to make merely one blowup. 
One first has to blow up the collision which leads to type $III$ singularity on the 
exceptional divisor. Singularities $III$ and $III^*$ are dual to each other and one has to make
another blowup introducing one extra component of the exceptional divisor.
Singularity $I^* _0$ corresponds to either $SO(8)$, $SO(7)$ or $G_2$,
depending on whether the singularity is split, supersplit, or nonsplit.
For simplicity we choose
the ``colliding'' group to be $G_2$.
The gauge group appears to be $G_2\times SU(2)\times E_7$ with the matter content
$(3n_1 +6) ({\bf 7,1,1})+ ({1\over 2}n_2+3)({\bf 1,1,56})$ and the mixed representations
\eqn\cont{
{1 \over 2}({\bf 7,2,1})+
{1 \over 2}({\bf 1,2,1})
}
In this case we get two extra tensor multiplets. It is instructive to check  the
dimension of the hypermultiplet moduli space (for simplicity we consider the
base being $P^1 \times P^1$). 
The number of independent parameters in polynomials $f, g$
is $7 \cdot 6+10 \cdot 8-4=118$. The total number of $n_H-n_V=273-29 n_T$.
Taking into account that we have two extra tensor  multiplets we get an identity
\eqn\trivial{118+ 6 \cdot 7 +7+1+ 3\cdot 56 -14-3-133=186~.}

In this example the full symmetry group is given by $G_2\times SU(2)\times E_7$. 
Each factor has its own coupling constant, governed by the area of the corresponding divisor
\area. Consider specific regime, when $G_2$ coupling approaches zero, keeping
$SU(2)$ coupling finite. In this regime we recover the  $SU(2)$ gauge theory with 
{\it global} $G_2$
symmetry similar to the examples of phase transitions discussed in \Wilast .

\bigskip

$\bullet$ $III^* \times III^*$ collision ($E_7 \times E_7$). 
In this case one needs to make five blowups.
The first one leads to a $I^*_0$ singularity on the exceptional divisor.
In turn, as it was discussed above,
the collision $III^*\times I^*_0$ requires two blowups of 
the base which produce additional components of the 
exceptional divisor with type $III$ and $I_0$ singularities.
The gauge group $E_7 \times E_7$ gets enhanced by a factor 
$SU(2)\times SO(7) \times SU(2)$, which follows from the structure of 
the singular locus.
The matter content is given by  
${1 \over 2}(n_i+5) {\bf 56}$ and 
\eqn\contIIIstar{ {1\over 2} ({\bf 1,2,8,1,1})+{1\over 2} ({\bf 1,1,8,2,1})
.}
Let us check the dimension of the hypermultiplet moduli space.
The number of independent parameters in the polynomials $f$ and $g$
is $6^2+8^2-4=96$.
Taking into account that we have two extra tensor  multiplets 
(i.e. $n_H-n_V=99$), we get an identity
\eqn\trivialll{96+ 2 \cdot 8 +2\cdot{5\over 2}\cdot
56 -2\cdot 3-21-2\cdot133=99~.}
It is worth mentioning that in both these examples we identify 
type  $III$ singularity with $SU(2)$ gauge group. 

Note that by giving expectation values to the scalar components of universal part of matter
one can higgs the gauge group on the exceptional divisor  down to
$SU(3)$ without any matter.
This $SU(3)$ group cannot be higgsed further.

\subsec{$J=0$ collisions}

The collisions $II\times I_0^*$, $II\times IV^*$ and $IV\times I_0^*$ 
do not require the blowup of the base. 
All other collisions require the blowup of the base, in some cases several times. 
In the table below we summarize 
the results on those collisions that can be resolved, preserving
the Calabi-Yau condition.

\bigskip
Table 7.  Gauge groups and matter for $J=0$ collisions.
\vskip .1in
\begintable
 Collision | Resolution | Gauge groups   \elt
I$^{*ns} _0 \times $IV$^{*ns}$ | I$^* _0$, II,  IV$^*$| $G_2 \times F_4$ \elt
 IV$^{*ns} \times $IV$^{*ns} $| IV$^*$, I$_0$, IV, I$_0$, IV$^*$ |   $F_4  \times SU(3)\times  F_4$
\endtable
\bigskip

$\bullet$ $I^* _0 \times IV^*$ collision.
In this case one need to make one blowup of the base.
For simplicity we consider a collision of  {\it generic} singularities, which corresponds to
$G_2 \times F_4$ gauge group. From the anomaly factorization conditions we get the following matter content
$(7+3n_1)({\bf 7,1})+ (4+n_2)({\bf  1,26})$ without any mixed representations.

The number of parameters in the polynomials $f$ and $g$ is equal to 128.
Therefore we have
\eqn\dimGF{128+49-14+4\cdot 26-52=215}
that coincides with $273-2\cdot 29$.
Note that this collision is related by the usual Higgs mechanism to 
$IV \times IV^*$ ($A_2 \times E_6$),
which leads to the $SU(3) \times F_4$ gauge group.   

$\bullet$ $IV^* \times IV^*$ collision.
In this case one needs to make three blowups of the base.
For simplicity we consider a collision of  {\it generic} singularities, which corresponds
to the
$F_4 \times F_4$ gauge group. This gauge group get enhanced by an extra 
$SU(3)$ factor. 
(The case of $E_6 \times E_6$ can be 
considered in a similar way.)
It is interesting that  type $IV$ singularity on one 
of the components of the exceptional divisor 
 corresponds to $SU(3)$ group {\it without matter}
and thus it  could not be higgsed down.
More specifically, from the anomaly factorization conditions we get the following matter content
$(3+n_1)({\bf 1,1,26})+(3+n_2)({\bf 26,1,1})$. 
Again, the dimension of the hypermultiplet moduli space 
\eqn\dimFSUF{113+6\cdot 26-2\cdot 52-8=157}
agrees with $n_H-n_V=273-4\cdot 29$.

\newsec{Tensionless strings and phase transitions}

We found that resolution of colliding singularities sometimes leads
 to enhanced gauge symmetry.  It is interesting to summarize possible gauge groups that
appear on the exceptional divisor. 
There is an infinite series of examples where the gauge group gets enhanced by 
$Sp((n+m)/2)$ factor with $(n+m+8)({\bf n+m})$ matter multiplets.
We also found $SU(2)$ with four doublets, 
$SU(3)$ without any matter  
and $SO(7)$  with two spinors. 

When the area of the blown up sphere tends to zero the 
corresponding gauge theory approaches the strong coupling regime.
It appears that  $Sp((n+m)/2)$ gauge theory with $(n+m+8)({\bf n+m})$ matter multiplets
 is in the same universality class as  F-theory compactification  on elliptic 
Calabi-Yau manifold  with the base being ${\bf F}_1$.  This theory exhibits phase transition which is due to  small $E_8$ instantons (see ref. \MV).  In fact, all resolutions 
with one blowup  ($\delta n_T=1$) discussed in this paper are in the same universality class.
They can be continuously deformed into each other by adjusting some 
hypermultiplets (higgsing  and unhiggsing).
In this process one can  break the gauge groups completely,
keeping the possibility of making blow (-up or -down) intact.
In this case we end up with the point-like $E_8$ instanton without
any gauge group on top of it.

The other phase transitions are new\foot{We are grateful to Cumrun Vafa
for stimulating discussion on phase transitions and small instantons.}. Indeed the phase transition with $\delta n_T=2$ is very different
from the above ones.
It corresponds to $SU(2)$ gauge theory with four doublets
 ($I^*_0\times III^*$ collision, two blowups). 
The spectrum formally coincides with that of F-theory compactification  on elliptic 
Calabi-Yau manifold  with the base being ${\bf F}_2$. This is consistent with the fact that
in case of ${\bf F}_2$ it is known that one cannot go to a new phase by making one blowup (blowdown).
The careful analysis of the elliptic fibration implies at at the very last stage of the higgsation  process we end up with $SU(2)$ gauge group with four doublets. The dimension of the Higgs branch is $8-3=5$.
At the same time
we have only $4$ parameters 
in the elliptic fibration to relax in order to break $SU(2)$, keeping the possibility of making two blowups on top of each other. One extra parameter appears when we relax the condition that  two blowup are on top of each other.  This observation is consistent with the fact that two Calabi-Yau spaces,  one elliptically fibered over the base with two blowup at {\it different points} in the base and the other elliptically fibered over the base with two blowup 
{\it on top of each other} are related by complex deformations.  Therefore this phase transition can be described by two small  $E_8$ instantons either at separate points or on top of each other depending on whether we keep $SU(2)$ gauge group broken or unbroken.

The case of $SU(3)$ gauge theory   ($IV^{ns *}\times IV^{ns *}$ collision) requires three blowups. The gauge group $SU(3)$ is generic and cannot be higgsed away. 
The remaining case of  $SU(2)\times SO(7) \times SU(2)$ 
 ($III^{ns *}\times III^{ns *}$ collision) requires five blowups. 
The gauge group can be higgsed down to $SU(3)$.  We believe that last two cases represent  new universality classes of phase transitions.
The interpretation of these phase transitions in terms of small instantons  
deserves further studies.

It is also worth mentioning that in the case of several blowups we get tensionless strings of different types, each corresponding to 3-branes wrapped around vanishing 2-cycles.
The blown up cycles have a nontrivial intersection matrix which implies that  nearly tensionless strings interact with each other. That presumably means that 
tensor multiplets also have nontrivial coupling. 

It is interesting to note that in the process of blowing down one shrinks a 4-cycle
(fibration of the singular fiber over the blown down $P^1$, or collection of $P^1$s)
to a singular fiber.  It is tempting to suggest that this 4-cycle is related to generalized del Pezzo
surface, but the precise relation is unclear\foot{We are grateful to Sheldon Katz for
the communications on this point.}. 

\bigskip

So far we have analysed the collisions that can be resolved by blowing up the base without 
any violation of the Calabi-Yau condition.
As mentioned above, these collisions lead to the phase transitions 
to new branches with different numbers of tensor multiplets, 
and hence they are closely related to the dynamics of tensionless strings.
However, there are possible collisions of different type for which a resolution of 
the base would violate the Calabi-Yau condition.
Here, we discuss some aspects of these collisions.
For convenience we summarize possible collisions in the table below. 
How many blowups should be done is 
indicated by the number. The sign ``$-$'' means that the blowup procedure
violates Calabi-Yau condition.

\bigskip
\vbox{Table 8. Possible collisions.
\vskip .1in
\begintable
  | $I_0 ^*$ | $I^* _{n>0}$ | $I_n$ | $II^*$ | $II$ | $IV^*$ | $IV$ | $III^*$ | $III$ \elt
 $I_0 ^*, ~J={\rm any}$ | $1$ | $1$ | $0$ | $4-$ | $0$ | $1$ | $0$ | $2$ | $0$ \elt
 $I^* _{n>0},~J= \infty$ | $1$ | $1$ | $0$ |  |  |  |  |  |  \elt
 $I_n,~J= \infty$ | $0$ | $0$ | $0$ |  |  |  |  |  |  \elt
 $II^*,~J= 0$ | $4-$ |  |  | $13-$ | $1$ | $6-$ | $3-$ |  |  \elt
 $II,~J= 0$ | $0$ |  |  | $1$ | $3-$ | $0$ | $1-$ |  |  \elt
 $IV^*,~J= 0$ | $1$ |  |  | $6-$ | $0$ | $3$ | $1$ |  |  \elt
 $IV,~J= 0$ | $0$ |  |  | $3-$ | $1-$ | $1$ | $3-$ |  |  \elt
 $III^*,~J= 1$ | $2$ |  |  |  |  |  |  | $5$ | $1$ \elt
 $III,~J= 1$ | $0$ |  |  |  |  |  |  | $1$ | $1$ 
\endtable}
\bigskip

It is instructive to consider the scheme of resolutions of collisions which violate 
the Calabi-Yau condition

$\bullet$ $I_0^*\times II^*\to I_0^*,IV,I_0^*,II,I_0, II^*$

$\bullet$ $IV^*\times II^*\to IV^*,II,I_0^*,IV,I_0^*,II,I_0, II^*$

$\bullet$ $II^*\times II^*\to II^*,I_0,II,I_0^*,IV,I_0^*II,IV^*,II,I_0^*,IV,I_0^*,II,I_0, II^*$

For simplicity we do not consider collisions which involve exceptional 
singularities $II,III,IV$.
It is curious to note that in all above three cases, 
the Calabi-Yau condition is violated at the step where one has to resolve
the $II \times IV$ collision, which cannot be avoided upon the resolution.
The physical meaning of the singularities $II$ and $IV$ in this situation is unclear,
even so we have assigned $SU(3)$ gauge group to $IV$ and nothing to $II$ 
when they appear in the collisions discussed in the previous section. 
The $II \times IV$ collision
corresponds to highly 
singular Calabi-Yau and does not lead to any Coulomb branch similar to those 
discovered in this paper.
This suggests that the nonperturbative dynamics at this  collision 
is not exhausted by tensionless strings and may imply an appearance of new physics.

On the other hand it seems that, say, in $II^*\times II^*$ collision,
it is not necessary to go through the whole chain of prescribed resolutions.
Let us suppose, for example, that we make just one first resolution.
In this case we get the collision $II^*\times IV^*\times II^*$.
In general this collision requires further resolutions of the base.
Instead we may try to smooth out $IV^*$ singularity
by complex deformations.
Thus we described a phase transition to a new branch with one new tensor multiplet.
Again, this transition is governed by tensionless strings.
However, in this case we are not able to describe the physics of the phase transition 
in terms of nonanomalous local field theory.

 A similar procedure of smoothing and singularizing elliptic fibration by complex deformations can be used to make a number of consequent phase transitions.

\newsec{Acknowledgments}

We are grateful to T. Pantev, V. Sadov and C. Vafa 
for useful discussions.
We also thank S. Katz and D. Morrison for helpful correspondence.
This work is partially supported by the
NSF grant PHY-92-18167, an NSF 1994 NYI award,  DOE 1994 OJI
award and Packard Foundation.

\listrefs
\end

We summarized the results on $I^* _n \times I^* _m$ collisions in the table below.

\bigskip
\begintable
 Collision | Gauge group | Matter ({\it universal} part) \elt
 $I^{*s} _2 \times I^{*s} _2$ | $SO(12) \times SO(5) \times SO(12)$ | ${1 \over 2}({\bf12,4,1})+  {1 \over 2}({\bf 1,4,12})$ \elt
 $I^{*s} _1 \times I^{*s} _2$ | $SO(10) \times SO(5) \times SO(12)$ | ${1 \over 2}({\bf10,4,1})+ ({\bf 1,4,1})+ {1 \over 2}({\bf 1,4,12})$ \elt
 $I^{*s} _1 \times I^{*s} _1$ | $SO(10) \times SU(2) \times SO(10)$ | $({\bf10,4,1})+ ({\bf 1,4,10})$ \elt
 $I^{*ss}_0 \times I^{*ss} _0$ | $SO(8)  \times SO(8)$ | none 
\endtable

\bigskip

It is interesting to note that all above discussed examples  with one extra 
tensor multiplet   
describe the same string phase.
They can be continuously deformed into each other by adjusting some 
hypermultiplets (higgsing  and unhiggsing).  This follows from the 
explicit form \WST. 
Moreover, In this process one can  break the gauge groups completely,
keeping the possibility of making blow (up or down) intact.
In this case we end up with the point-like instanton without
any gauge group on top of it.

  It is easy to count the codimension of such an event.
The polynomials $f(z,w)$ and $g(z,w)$ have the following expansions 
(see \WST)
\eqn\pol{f(z,w)=\sum f_{k,l} z^k w^l~,~~~g(z,w)=\sum g_{k,l} z^k w^l~,}
where coefficients $f_{k,l}$ differ from zero only if $k+l \geq 4$ and $k+2l \geq 8$; 
coefficients $g_{k,l}$ differ from zero only if $k+l \geq 6$ and $k+2l \geq 12$.
The codimension of two blowups on top of each other 
is equal to $62=2+29+2+29$, where we explicitly singled out the degrees of freedom corresponding to the positions of the instantons. This is precisely what one expects for two  $E_8$ point-like instantons.

The case of $SU(3)$ gauge theory   ($IV^{ns *}\times IV^{ns *}$ collision) requires three blowups. The gauge group $SU(3)$ is generic and cannot be higgsed away.
In the expansion \pol\ coefficients $f_{k,l}$ differ from zero only if $k+l \geq 4$,  $k+2l \geq 8$ and  $2k+l \geq 8$;
coefficients $g_{k,l}$ differ from zero only if $k+l \geq 6$, $k+2l \geq 12$ and $2k+l \geq 12$. That implies that one has to adjust $83$ parameters. 
The remaining case of  $SU(2)\times SO(7) \times SU(2)$ 
 ($III^{ns *}\times III^{ns *}$ collision) requires five blowups and $140$
parameters to adjust. 
The gauge group can be higgsed down to $SU(3)$.  We believe that last two cases represent  new universality classes of phase transitions.
The interpretation of these phase transitions in terms of small instantons is unclear and they deserve further studies. 

Indeed, the transition discovered here requires  two blowups.
The codimension counting suggests that this phase transition is related two 
two zero size $E_8$ instantons on top of each other. Indeed, in order to 
be able to make two blowup at the {\it same} point the polynomials $f(z,w)$ and 
$g(z,w)$ should have the following expansions 
(see \WST)
\eqn\pol{f(z,w)=\sum f_{k,l} z^k w^l~,~~~g(z,w)=\sum g_{k,l} z^k w^l~,}
where coefficients $f_{k,l}$ differ from zero only if $k+l \geq 4$ and $k+2l \geq 8$; 
coefficients $g_{k,l}$ differ from zero only if $k+l \geq 6$ and $k+2l \geq 12$.
Therefore one should adjust $59+1+2$ complex parameters, 
where we explicitly
singled out the location of the first blowup and the location of 
the second blowup on the blown up sphere.
Therefore, in this event one should adjust $59$ hypermultiplets, which is consistent with small instanton interpretation.
Indeed, putting two instantons on top of each other requires to adjust $29+29+1$ 
hypermutiplets, where one stands for adjusting
relative position of two instantons.  Naively, this argument seems to contradict the
anomaly relation which implies that the change in the number of hypermultiplets
should be equal to $58$. 
To break $SU(2)$ completly we 
The resolution of this paradox is that in the higgsation process
the blowup ceased to be on top of each other.